\documentclass{article}

\usepackage{PRIMEarxiv}

\usepackage[utf8]{inputenc} 
\usepackage[T1]{fontenc}    
\usepackage{hyperref}       
\usepackage{url}            
\usepackage{booktabs}       
\usepackage{amsfonts}       
\usepackage{nicefrac}       
\usepackage{microtype}      
\usepackage{lipsum}
\usepackage{fancyhdr}       
\usepackage{graphicx}       
\graphicspath{{media/}}     
\usepackage{cases}
\usepackage{subcaption}
\usepackage{listings}
\usepackage{color}

\definecolor{mygreen}{rgb}{0,0.6,0}
\definecolor{mygray}{rgb}{0.5,0.5,0.5}
\definecolor{mymauve}{rgb}{0.58,0,0.82}

\usepackage{graphicx}
\usepackage{float}
\usepackage{diagbox}
\usepackage{multirow}
\usepackage{makecell}

\title{\textit{Vignat}: \underline{V}ulnerability \underline{I}dentification by Learning Code Semantics via \underline{G}raph \underline{A}ttention Ne\underline{T}works
}

\author{
  Shuo Liu \\
  Columbia University\\
  New York, NY\\
  \texttt{sl4921@columbia.edu} \\
   \And
  Gail Kaiser \\
  Columbia University\\
  New York, NY\\
  \texttt{kaiser@cs.columbia.edu} \\
}

\begin{document}
\maketitle

\begin{abstract}
Vulnerability identification is crucial to protect software systems from attacks for cyber-security. However, huge projects have more than millions of lines of code, and the complex dependencies make it hard to carry out traditional static and dynamic methods. Furthermore, the semantic structure of various types of vulnerabilities differs greatly and may occur simultaneously, making general rule-based methods difficult to extend. In this paper, we propose \textit{Vignat}, a novel attention-based framework for identifying vulnerabilities by learning graph-level semantic representations of code. We represent codes with code property graphs (CPGs) in fine grain and use graph attention networks (GATs) for vulnerability detection. The results show that \textit{Vignat} is able to achieve $57.38\%$ accuracy on reliable datasets derived from popular C libraries. Furthermore, the interpretability of our GATs provides valuable insights into vulnerability patterns.
\end{abstract}

\keywords{Vunerability identification \and Graph attention networks \and Code property graph}

\section{Introduction}
Vulnerability identification is crucial in ensuring the security, integrity, and resilience of systems, networks, and applications. By proactively identifying and addressing vulnerabilities, organizations can reduce the risk of security breaches, disruption of services and malware propagation. Vulnerability identification has promising applications in fields such as software development, where it can be used to improve code quality and reduce the likelihood of introducing vulnerabilities in the first place, and in security research, where it can be used to develop new defensive technologies and techniques.

Some works identify vulnerabilities by executing testing programs and analyzing runtime behaviors, such as \cite{sutton2007fuzzing}, \cite{cadar2008klee}, and \cite{stephens2016driller}. However, due to the limited code coverage, execution dependencies, and substantial overhead associated with dynamic approaches, researchers start to focus on more efficient static vulnerability detection methods. There are several challenges that make static vulnerability detection difficult. Since large-scale projects often consist of millions of lines and exhibit complex calling relationships within and between projects, it is hard to effectively feature and analyze the code. Moreover, real-world vulnerabilities are sparse, which can cause bias when confronted with new and unseen patterns.

Some rule-based static analyzers have been developed to locate vulnerabilities using defined patterns or signatures that match known vulnerabilities or programming errors, \cite{lawall2008coccinelle}, \cite{clang2019analyzer}, and \cite{wheeler2021flawfinder}. While these methods are useful in detecting common types of vulnerabilities, they heavily rely on predefined rules and struggle to handle complex projects, making them unsatisfactory in real-world scenarios. Besides, some transformer-based models are also applied to this task because of their sequence processing ability and interpretability, \cite{BERT}, \cite{Sanh2019DistilBERTAD}, \cite{RoBERTA}, and \cite{feng2020codebert}. Nonetheless, these studies are difficult to obtain a comprehensive understanding of program semantics solely. Because, unlike natural language sequences, source codes are more structured, and the vulnerabilities can often be subtle flaws that need thorough examination from various semantic perspectives. Therefore, it seems a more reasonable way to find the patterns of vulnerabilities by analyzing the complex relationships in various code representations.

In this paper, we address these challenges by constructing graph embeddings of code functions and analyzing the logical relationships between tokens to predict code vulnerability. Using the self-attention mechanism \cite{vaswani2017attention}, our models exhibit excellent explainability, enabling us to infer vulnerability patterns instead of relying on experience to detect vulnerabilities. Our main contributions are: 

\begin{itemize}
    \item We obtain a comprehensive representation of code functions by embedding them into CPGs to capture code syntactic structure, control flow, and data dependencies.
    \item We propose a GAT framework, \textit{Vignat}, for capturing and modeling complex relationships among nodes in CPGs, where higher attention edges can reveal patterns of vulnerabilities.
    \item We evaluate the effectiveness of \textit{Vignat} on manually labeled datasets collected from 4 large-scale C projects. \textit{Vignat} achieves better results on the dataset with various attention-based models, up to 10\% accuracy and 5\% F-score improvement over baseline methods.
\end{itemize}

\begin{figure*}[t]
    \centering
    \includegraphics[scale=.3]{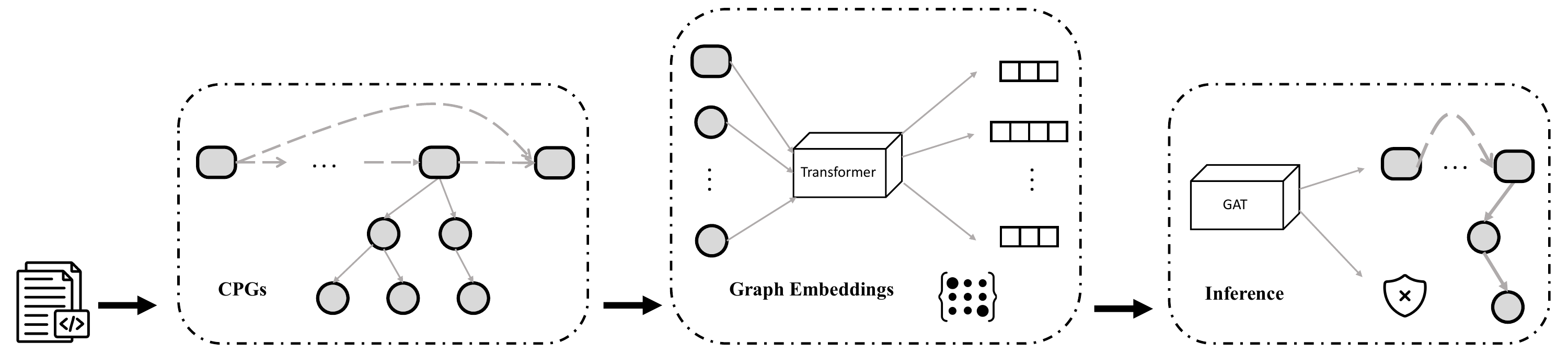}
    \caption{An overview of \textit{Vignat} framework.}
    \label{fig:framework}
\end{figure*}

The rest of this paper is organized as follows. Section \ref{sec:framework} presents \textit{Vignat} framework. Section \ref{sec:evaluation} demonstrates the performance and illustrates the interpretability of our model. Section \ref{sec:conclu} concludes this paper.

\section{\textit{Vignat} Framework} \label{sec:framework}

This section provides an overview of our \textit{Vignat} framework, as shown in Fig. \ref{fig:framework}. Within the \textit{Vignat} framework, we first tokenize the source code functions and construct CPGs, a composite representation of code semantics. Following this, we employ various embedding methods to obtain node embeddings. In conjunction with graph connectivity, the graphs are inputted into a GAT for graph-level predictions. By extracting attention weights from the output of the attention layer, we can identify salient edges, revealing patterns associated with vulnerabilities.

\subsection{Graph Embedding of Code}

\begin{figure*}[htb]
    \centering
    \begin{subfigure}[b]{0.24\textwidth}
        \centering
        \begin{lstlisting}[language=C]
void func() {    
  int x = source();    
  if (isEven(x)) {
    proceed(10 / x);
  }
}\end{lstlisting}
    \end{subfigure}
    \hfill
    \begin{subfigure}[b]{0.72\textwidth}
        \centering
        \includegraphics[width=\textwidth]{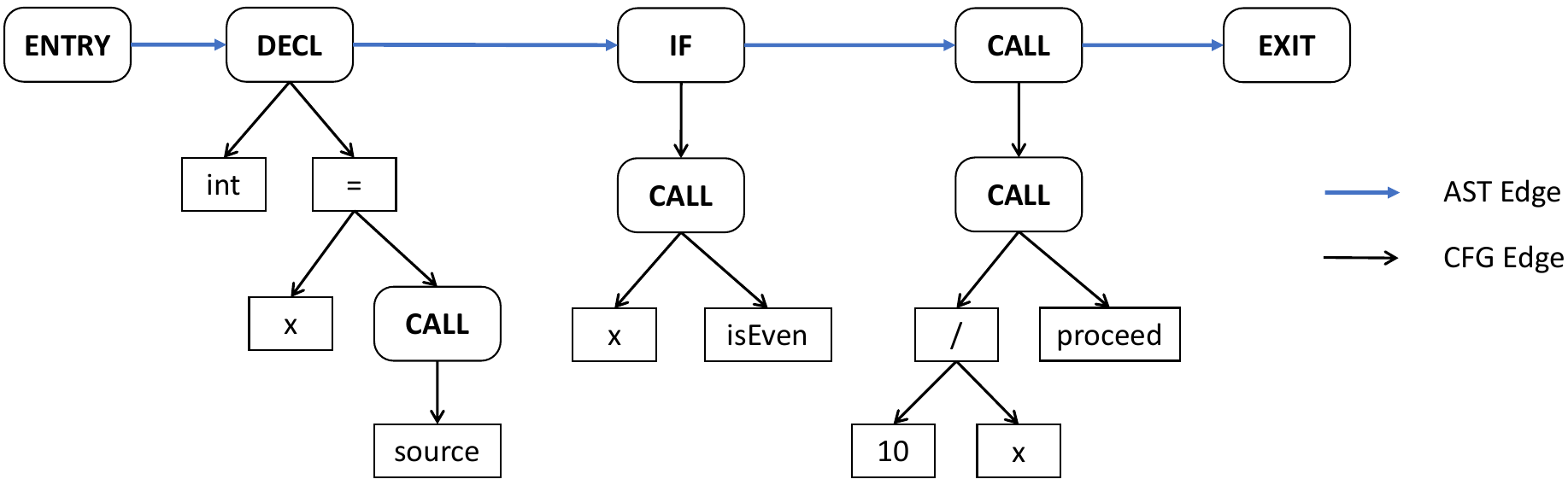}
    \end{subfigure}
    \caption{A vulnerable code snippet and its corresponding CPG.}
    \label{fig:cpg}
\end{figure*}

Aside from the semantics conveyed by code tokens and the logic present in natural code sequencing (NCS), highly structured graph representations of code obscure a significant amount of logical information. The abstract syntax tree (AST) is a tree-like representation that encodes how statements and expressions are nested to produce a program. Inner nodes denote operators, leaf nodes denote operands, and edges specify container and content relationships. The control flow graph (CFG) describes the order in which code statements are executed. It also shows the conditions that must be met for a particular execution path. In a CFG, nodes represent statements or predicates, and edges denote the paths the program can traverse. The program dependence graph (PDG) represents dependencies among statements and predicates in a program. Data dependence edges show how a node's outcome impacts a variable in another node, while control dependence edges reveal the effect of predicates on variable values. Integrating elements of ASTs, CFGs, PDGs, a CPG provides a unified representation for program analysis and enables us to simultaneously reason about all perspectives of code properties. A CPG for a vulnerable code snippet is shown in Fig. \ref{fig:cpg}.

In this paper, we make use of an open-source code analysis platform \texttt{Joern} to parse source code functions and construct CPGs in batches. In a CPG, nodes represent program constructs, including variables, methods, control structure, etc. Each node contains several tokens and has some attributes according to its type, such as the name of local variables, the signature of the method, and the type of control structure. Nodes are connected by directed edges to represent relationships in the program structure. In \textit{Vignat}, Considering that a node may have different numbers of tokens in it, we obtain its embedding by averaging its token embeddings. To make the graph embedding tractable for graph neural networks (GNNs), we simplify the CPGs by disregarding the heterogeneity of various relationships and eliminating duplicate edges. Only the connectivity information is passed to the subsequent GAT model.

\subsection{GAT}

Graph Neural Networks (GNNs) have emerged as a family of models for learning representations of graph-structured data. A graph $\mathcal{G} = (\mathcal{V}, \mathcal{E})$ consists of a set of nodes $\mathcal{V}$ and edges $\mathcal{E}$, where each edge $(i, j)$ represents a connection between nodes $i$ and $j$. The key idea behind GNNs is to iteratively update node representations by aggregating information from neighboring nodes. Let $\mathbf{X}$ be the initial node feature matrix, where each row corresponds to the feature vector of a node. Mathematically, the propagation of a GNN can be expressed as:

\begin{equation}
    \mathbf{h}^{(l)} = \sigma\left(\mathbf{A} \cdot ReLU\left(\mathbf{W}^{(l-1)} \cdot \mathbf{h}^{(l-1)}\right)\right)
\end{equation}

where $\mathbf{h}^{(l)}$ is the node representation matrix at layer $l$, $\mathbf{W}^{(l)}$ is the learnable weight matrix, $\mathbf{A}$ is the adjacency matrix capturing graph connectivity, and $\sigma$ denotes an activation function. This layer-wise aggregation process allows GNNs to capture complex relationships in graph-structured data and learn expressive node representations.

GAT was introduced by \cite{velickovic2018graph} as a novel approach for graph representation learning, which can naturally generalize convolutions to irregular graph structures. GAT incorporates self-attention mechanisms, which enables the model to weigh the importance of neighboring nodes for the target one, thus capturing the local code structure of the CPG.

The GAT model consists of a series of graph attention layers. Each layer computes a new set of node embeddings based on the input node features and the learned attention coefficients. The attention coefficients between nodes $i$ and $j$ in layer $l$ can be computed as follows:
\begin{equation}
    e_{ij}^{(l)} = LeakyReLU\left(\textbf{a}^{(l)T}\left[\textbf{W}^{(l)}\textbf{h}_i^{(l-1)} \parallel \textbf{W}^{(l)}\textbf{h}_j^{(l-1)}\right]\right),
\end{equation}

where $\parallel$ denotes the concatenation operation, and $\textbf{a}^{(l)}$ is a learnable vector.

To ensure that the coefficients are normalized across the neighboring nodes, we apply the softmax function:
\begin{equation}
    \alpha_{ij}^{(l)} = softmax_j\left(e_{ij}^{(l)}\right) = \frac{exp(e_{ij}^{(l)})}{\sum_{k \in \mathcal{N}(i)}^{} exp(e_{ik}^{(l)})},
\end{equation}

where $\mathcal{N}(i)$ denotes the set of neighboring nodes of node $i$. The new node embeddings for layer $l$ are computed as a weighted sum of the input features of the neighboring nodes:
\begin{equation}
    \textbf{h}_i^{(l)} = \sigma\left(\sum_{j \in \mathcal{N}(i)}^{} \alpha_{ij}^{(l)}\textbf{W}^{(l)}\textbf{h}_j^{(l-1)}\right),
\end{equation}

where $\sigma$ is an activation function, such as the rectified linear unit (ReLU).

One of the key advantages of GAT is its explainability. In contrast to graph convolutional networks (GCNs) \cite{kipf2016semi}, where the aggregation of neighboring node features is often performed uniformly, making it difficult to discern the influence of specific nodes, GAT provides an interpretable measure of the importance of neighboring nodes for a target node. By investigating the attention coefficients $e_{ij}^{(l)}$, we are able to gain insights into the rationale of our model, which is useful for identifying influential paths and addressing potential biases.
 
\section{Experiments} \label{sec:evaluation}

We evaluate \textit{Vignat} with the following research questions.

\begin{enumerate}
    \item How does \textit{Vignat} compare to the transformers that are based on the NCS?
    \item How does \textit{Vignat} perform on different kinds of code representations?
    \item How do different kinds of word embedding methods affect the performance of \textit{Vignat}?
    \item How does attention-mechanism powered \textit{Vignat} compare to GCNs?
    \item How to infer patterns of code vulnerabilities through attention coefficients?
\end{enumerate}

\subsection{Setup}

We use the \textit{Devign} dataset to evaluate our models \cite{zhou2019devign}. Compared to datasets with labels generated by static analyzers and artificial ones, it is more reliable and maintains a sparsity of vulnerabilities and is more reliable by manually labeling vulnerabilities in large-scale, real-world C projects. Specifically, we focused on two projects, \texttt{FFmpeg} and \texttt{QEMU}, and selected medium-length functions containing less than $1200$ tokens as samples for model training and testing. As the DFG edges are labeled with the variables involved, it tremendously complicates embedded graphs. Therefore, we only extracted ASTs and CFGs and combined them into composite graph as CPGs for our dataset using \texttt{Joern}. We calculate the average embedding of tokens within a node to obtain its embedding and standardize its embedding size to $768$ using pretrained models, BERT, DistilBERT, RoBERTa, and CodeBERT. If a CPG has more than $225$ nodes, we cut the first $225$ ones; otherwise, we use padding zeros to keep the size of input code features same.

We configure the GAT models as follows. Models are configured with the input hidden dimension of $225$ and learning rate $0.0001$. We implement our models using PyTorch 2.0.0, CUDA 11.8, and train them $100$ epochs with batch size $8$ using 16GB Tesla V100-SXM2 GPU. It takes us $0.0358$ seconds to generate each CPG in average. And we takes $116$ minutes to train For \textit{Vignat} with Roberta embedding on \texttt{FFmpeg}, we spend $116$ minutes to train; It takes us $127$ minutes to train \textit{Vignat} with Roberta embedding on \texttt{QEMU}.

\subsection{Result Analysis}

To evaluate the performance of \textit{Vignat}, we compare it with some baseline methods, including state-of-the-art transformers based on the NCS and GCNs based on graph representation of codes. To avoid the resulting gap caused by data heterogeneity, we provide these baselines with the same training set and test set after shuffle and split. We measure the performance of different methods from 4 metrics, accuracy, precision, recall, and F1 score. The comparison results are shown in Table. \ref{table:compare}.

\renewcommand{\arraystretch}{1.3}
\begin{table*}[ht]
{
\scriptsize
	\centering
	\begin{tabular}{c|c|c|cccc}
		\hline \hline
            \multirow{2}{*}{\makecell{Class}} & \multirow{2}{*}{\makecell{Embedding}} & \multirow{2}{*}{\makecell{Representation}} &  \multicolumn{4}{c}{FFmpeg}  \\
            \cline{4-7} & & & Acc & Prec & Rec & F1 \\
            \hline
            \multirow{4}{*}{Transformer} & BERT & \multirow{4}{*}{NCS} & $53.33\%$ & $55.77\%$ & $90.63\%$ & $68.89\%$
            \\
             & DistilBERT & & $56.67\%$ & $54.45\%$ & $96.88\%$ & $69.05\%$
             \\
             & RoBERTa & & $53.33\%$ & $53.85\%$ & $87.50\%$ & $66.67\%$ 
              \\
             & CodeBERT & & $51.67\%$ & $52.83\%$ & $87.50\%$ & $65.88\%$
             \\
             \hline
             \multirow{15}{*}{GCN} &\multirow{3}{*}{Word2Vec} & AST & $54.09\%$ & $53.70\%$ & $90.63\%$ & $67.44\%$ 
             \\
              & & CFG &$52.46\%$ & $52.46\%$ & $100.00\%$ & $68.82\%$ 
              \\
              & & CPG & $55.74\%$ & $56.10\%$ & $71.88\%$ & $63.10\%$ 
              \\
              \cline{2-7} &\multirow{3}{*}{BERT} & AST & $50.82\%$ & $52.00\%$ & $81.25\%$ & $63.41\%$
              \\
              & & CFG & $52.46\%$ & $52.46\%$ & $100.00\%$ & $68.82\%$
              \\
              & & CPG & $57.38\%$ & $56.82\%$ & $78.13\%$ & $65.79\%$
              \\
              \cline{2-7} &\multirow{3}{*}{DistilBERT} & AST & $50.82\%$ & $51.92\%$ & $84.37\%$ & $64.29\%$ 
              \\
              & & CFG & $54.10\%$ & $53.57\%$ & $93.75\%$ & $68.18\%$
            \\
              & & CPG & $50.82\%$ & $51.92\%$ & $84.38\%$ & $64.29\%$
             \\
              \cline{2-7} &\multirow{3}{*}{\textbf{RoBERTa}} & AST & $50.82\%$ & $51.92\%$ & $84.37\%$ & $64.29\%$ 
              \\
              & & CFG & $52.46\%$ & $52.46\%$ & $100.00\%$ & $68.82\%$
             \\
              & & \textbf{CPG} & $\textbf{57.38\%}$ & $\textbf{56.52\%}$ & $\textbf{81.25\%}$ & $\textbf{66.67\%}$
              \\
              \cline{2-7} &\multirow{3}{*}{CodeBERT} & AST & $52.46\%$ & $52.46\%$ & $56.25\%$ & $55.38\%$
              \\
              & & CFG & $52.46\%$ & $52.46\%$ & $100.00\%$ & $68.82\%$
              \\
              & & CPG & $57.38\%$ & $57.50\%$ & $71.88\%$ & $63.89\%$ 
              \\
              \hline
             \multirow{6}{*}{\textbf{GAT}} &\multirow{3}{*}{Word2Vec} & AST & $55.74\%$ & $56.10\%$ & $71.86\%$ & $63.01\%$
              \\
              & & CFG &$52.46\%$ & $52.46\%$ & $100.00\%$ & $68.82\%$ 
               \\
              & & \textbf{CPG} & $\textbf{63.93\%}$ & $\textbf{63.16\%}$ & $\textbf{75.00\%}$ & $\textbf{68.58\%}$ 
              \\
              \cline{2-7} &\multirow{3}{*}{BERT} & AST & $59.02\%$ & $57.14\%$ & $87.50\%$ & $69.14\%$ \\
              & & CFG & $52.46\%$ & $52.46\%$ & $100.00\%$ & $68.82\%$
              \\
              & & CPG & $60.66\%$ & $59.09\%$ & $81.25\%$ & $68.42\%$ \\
              \hline \hline
	\end{tabular}
	\caption{Comparison of different methods. Acc, Prec, Rec, F1 represent accuracy, precision, recall, and   F1 score respectively.}
	\label{table:compare}}
\end{table*}

As shown in Table. \ref{table:compare}, on \texttt{FFmpeg} dataset, \textit{Vignat} has the best performance over all models. Compared with transformer baselines, it improves $4.05\%$ accuracy over BERT and RoBERTa, and $5.71\%$ over CodeBERT. We find that CPGs provide the most comprehensive representation among the graphs, as the accuracy improves $6.56\%$ over ASTs, and $4.92\%$ over CFGs. Besides, experiments also show that different embedding methods have an impact on model performance. RoBERTa performs well on all data sets. For \textit{Vignat} using GCNs, though it does not show a substantial enhancement compared to alternative embedding techniques on accuracy, it has a better overall performance with higher precision and recall.

\subsection{Pattern Explanation}

\begin{figure*}[htb]
    \centering
    \includegraphics[scale=.5]{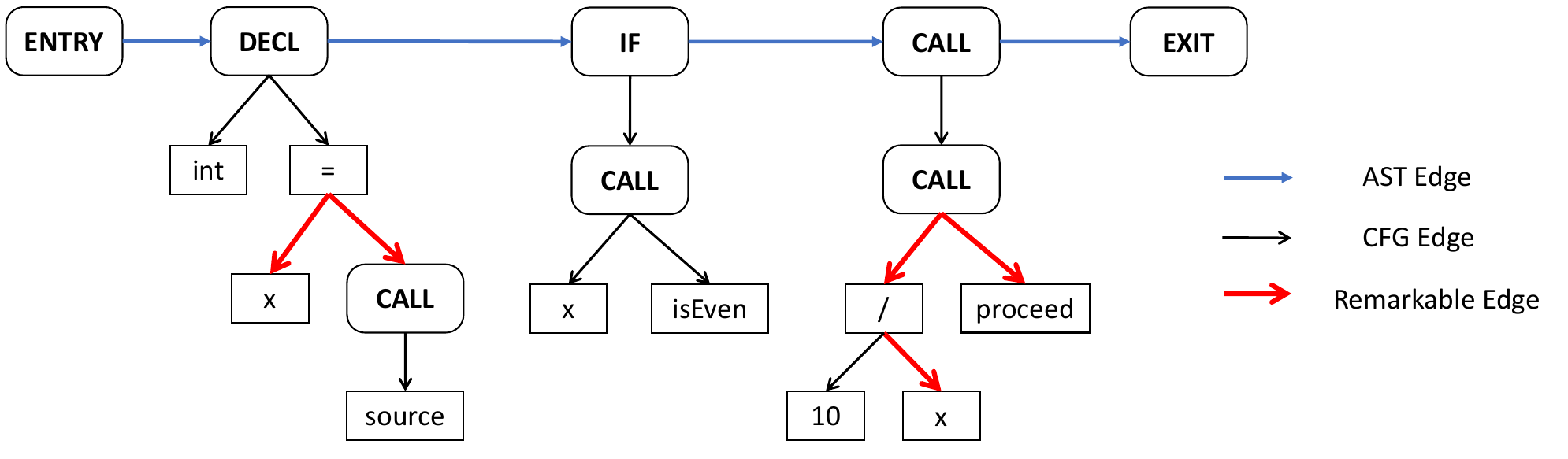}
    \caption{5 edges with highest attention in Fig. \ref{fig:cpg}.}
    \label{fig:cpg_marked}
\end{figure*}

Fig. \ref{fig:cpg_marked} illustrates 5 CPG edges with the largest values of attention coefficients $e^{(l)}_{ij}$ for the source code in Fig. \ref{fig:cpg}. In this code example predicted to be vulnerable, the type and operator of a variable are highlighted with red arrows. These edges reveal the vulnerabilities that could occur in this code snippet. Specifically, the integer may overflow when declaring the variable and the divisor can be zero, and these vulnerabilities will affect the following proceeding functions.

\section{Conclusion}
\label{sec:conclu}

This paper presents \textit{Vignat}, an innovative attention-based framework for vulnerability identification. By leveraging CPGs to comprehensively capture the syntactic structure, control flow, and data dependencies of code functions, and employing GATs to model complex relationships among nodes, \textit{Vignat} effectively predicts and reasons the vulnerabilities in source codes. Our method demonstrates remarkable performance, achieving up to $57.38\%$ accuracy on reliable datasets, and at least $4\%$ accuracy improvement over transformers and $1\%$-$5\%$ over other GNNs. Furthermore, by employing the interpretability of GATs in \textit{Vignat}, we are able to get valuable insights into vulnerability patterns, paving the way for future advances in vulnerability identification and prevention in cybersecurity.

\section{Acknowledgements}

Kaiser is supported in part by DARPA/NIWC-Pacific N66001-21-C-4018 and in part by NSF CNS–2247370 and CCF-2313055.

\bibliographystyle{unsrt}  
\bibliography{references}

\end{document}